\documentclass[a4paper,12pt,notitlepage]{article}

\usepackage{amsmath}
\usepackage{latexsym}
\usepackage{amssymb}
\usepackage{amsfonts}
\usepackage[dvipdfm]{color}

\numberwithin{equation}{section}

\def\colon{{:}\;}

\newcommand{\R}{ \mathbb{R} }
\newcommand{\C}{ \mathbb{C} }
\newcommand{\re}{\mathop{\mathrm{Re}}}
\newcommand{\im}{\mathop{\mathrm{Im}}}
\newcommand{\tr}{\mathop{\mathrm{tr}}}
\newcommand{\diag}{\mathop{\mathrm{diag}}}

\newtheorem{Thm}{Theorem}[section]
\newtheorem{Lemma}[Thm]{Lemma}
\newtheorem{Prop}[Thm]{Proposition}

\author{Ilya Ya. Goldsheid$^{1}$ and Boris A.
Khoruzhenko$^{1,2}$  \vspace{1ex}\and$^{1}$School of Mathematical
Sciences, Queen Mary, University of London,\\
London E1 4NS, U.K.
\and $^{2}$Department of Mathematical Sciences, Brunel
University, \\ London UB8 3PH, U.K.}

\title{Thouless formula for random non-Hermitian Jacobi matrices}

\begin{document}

\maketitle

\begin{abstract}
Random non-Hermitian Jacobi matrices $J_n$ of increasing dimension
$n$ are considered. We prove that the normalized eigenvalue
counting measure of $J_n$ converges weakly to a limiting measure
$\mu$ as $n\to\infty$. We also extend to the non-Hermitian case
the Thouless formula relating $\mu$ and the Lyapunov exponent of
the second-order difference equation associated with the sequence
$J_n$. The measure $\mu$ is shown to be log-H\"older continuous.
\end{abstract}

\section{Introduction}\label{sec1}
Let $a_j$, $b_j$, and $c_j$ are three given sequences of complex
numbers. Consider the second-order difference equation for $f$
\begin{equation}\label{1}
a_jf_{j-1}+ b_jf_j +c_{j}f_{j+1}=zf_j, \hspace{3ex} j=1,2, \ldots.
\end{equation}
This equation can be also written as
\begin{equation}\label{ggg}
 \left(
\begin{array}{l}
f_{j+1} \\ f_j \\
\end{array}
\right) =g_j \left(
\begin{array}{l}
f_{j} \\ f_{j-1} \\
\end{array}
\right),\ j=1,2, \ldots ,  \hspace{1ex} \hbox{ where }\ g_j=
\begin{pmatrix}
\frac{z-b_j}{c_j}  & \frac{-a_j}{c_j}   \\1   &  0    \\
\end{pmatrix}.
\end{equation}
Denote by $f_{j}(z)$ the solution of (\ref{1}) satisfying the
initial condition $f_0=0$, $f_1=1$. In terms of the transfer
matrix $S_n(z)=g_n\cdot \ldots \cdot g_1$,
\begin{equation}\label{3a}
 \left(
\begin{array}{l}
f_{n+1}(z) \\ f_n(z) \\
\end{array}
\right) =S_n(z) \left(
\begin{array}{l}
1 \\ 0 \\
\end{array}
\right).
\end{equation}
Obviously, $f_{n+1}(z)$ is a polynomial in $z$ of degree $n$,
\begin{equation}\label{f}
f_{n+1}(z)=k_n\prod_{l=1}^n (z-z_l), \hspace{3ex}
k_n=\prod_{j=1}^n 1/c_j.
\end{equation}
Its roots $z_1, \ldots z_n$ are the eigenvalues of the tridiagonal
(Jacobi) matrix
\begin{equation}\label{2}
J_n =
\begin{pmatrix}
  b_1 & c_1  &  &  &  \\[0.5ex]
  a_2  & b_2 & c_2 &  &  \\[0.5ex]
   & \ddots & \ddots & \ddots &  \\[0.5ex]
   &  & \hspace{2ex} a_{n-1}  & b_{n-1}& c_{n-1} \\[0.5ex]
   &   &  &\hspace{2ex}a_{n}  & b_n
\end{pmatrix}.
\end{equation}
In this paper we are concerned with the limiting distribution of
the eigenvalues of $J_n$ as $n\to\infty$ for random $a_j$, $b_j$,
and $c_j$.

If all $b_j$ are real and $c_j=a_{j+1}^*$ for all $j$ the matrices
$J_n$ are Hermitian. The eigenvalue distribution of such matrices
was studied extensively in the past in the context of the Anderson
model, see e.g. \cite{PF,CL}. In this case the eigenvalues are
always real and there are several ways to prove that the
normalized eigenvalue counting measure of $J_n$ converges to a
limiting measure as $n\to\infty$. None of these proofs works in
the non-Hermitian case and little is known about the limiting
eigenvalue distribution of random non-Hermitian Jacobi matrices,
however, see \cite{D,Zee}.

Our interest to such matrices is partly motivated by non-Hermitian
quantum mechanics of Hatano and Nelson \cite{HN1,HN2} which, in
one dimension, leads to equation (\ref{1}) with the coefficients
$a_j$, $ b_j$, and $c_j$ chosen randomly from the special class
defined by the restrictions
\begin{equation}\label{hn}
b_j \in \R \hspace{1ex} \hbox{and} \hspace{1ex} a^{*}_{j+1}/c_j>0
\hspace{1ex} \hbox{for all $j$.}
\end{equation}
In this class the Liouville
substitution\footnote{$f_j=\theta_j\psi_j$, where $\theta_1=1$ and
$\theta_k=(\prod_{j=1}^{k-1}a^*_{j+1}/c_j)^{1/2}$ for $k\ge 2$.}
reduces equation (\ref{1}) to the symmetric equation
\begin{equation}\label{symmetrized}
s_{j-1}^*\psi_{j-1} +b_j\psi_j+s_j\psi_{j=1}=z\psi_j
\end{equation}
where $s_j=c_j(a^*_{j+1}/c_j)^{1/2}$. However, the situation here
is much richer than in the Hermitian case as the choice of
boundary conditions to accompany equation (\ref{1}) has a profound
effect on the spectrum of the associated Jacobi matrix. If the
Dirichlet boundary conditions, $f_0=0$ and $f_{n+1}=0$, are chosen
then the corresponding Jacobi matrix is $J_n$ (\ref{2}). As the
Dirichlet boundary conditions are preserved by the Liouville
transformation, the spectrum of $J_n$ is real provided the
coefficients $(a_j,b_j,c_j)$ belong to the Hatano-Nelson class
(\ref{hn}). On the other hand, if one imposes the periodic
boundary conditions, $f_0=f_n$ and $f_1=f_{n+1}$, then the
spectrum of the corresponding Jacobi matrix turns out to be
complex. This is not surprising of course as the Liouville
substitution transforms the periodic boundary conditions for $f$
into highly asymmetric boundary conditions for $\psi$. What is
surprising however is that in the limit $n\to\infty$ the complex
eigenvalues lie on analytic curves \cite{GK1} and are regularly
spaced even if the coefficients in equation (\ref{1}) are chosen
randomly \cite{GK2}. These effects are specific to the
Hatano-Nelson class and the proofs and analysis of the limiting
eigenvalue distribution given in \cite{GK1,GK2} exploit the
relation between equations (\ref{1}) and (\ref{symmetrized}). Of
course, in the general case of arbitrary coefficients no such
relation exists and one requires a different approach in order to
investigate the eigenvalue distribution of $J_n$. We develop such
an approach in the present paper.

Throughout this paper we assume that:
\begin{itemize}
\item[{\bf A1}]$\{(a_j,b_j,c_j)\}_{j=1}^{\infty}$ is a sequence
i.i.d. random vectors.

\item[{\bf A2}] For some $\delta>0$
$E[|a_j|^{\delta}+|a_j|^{-\delta} +|b_j|^{\delta}
+|c_j|^{\delta}+|c_j|^{-\delta}]<\infty$.

\item[{\bf A3}] The support of the probability distribution
of the random vector $(a_1,b_1,c_1)$ contains at least two
different points $(a,b,c)$ and $(a^{\prime},b^{\prime},
c^{\prime})$.
\end{itemize}

If all mass of the probability distribution of $(a_j,b_j,c_j)$ is
concentrated at one point $(a,b,c)$ then of course we have a
tridiagonal matrix with constant diagonals. This is a particular
case of T\"oplitz matrices. Eigenvalue distribution of
non-Hermitian T\"oplitz matrices was extensively studied in the
past, see e.g. survey \cite{W2}.

Our main result expresses the limiting distribution of the
eigenvalues of $J_n$ in terms of the (upper) Lyapunov exponent
\[
\gamma (z)=\lim_{n\to\infty}\frac{1}{2n}\log [|f_{n+1}(z)|^2
+ |f_{n}(z)|^2]
\]
of equation (\ref{1}).
It is well known that (for every complex $z$) the above limit
exists with probability one and is nonrandom. This follows from
Oseledec's multiplicative ergodic theorem \cite{O}. A more subtle
fact is that in our case $\gamma (z)$ can be calculated using the
well known Furstenberg formula \cite{F}, and moreover
\begin{equation}\label{1a}
\gamma (z)=\lim_{n\to\infty}\frac{1}{n} E \log ||S_n(z)||,
\hspace{3ex} z\in \C.
\end{equation}
The function $\gamma (z)$ is
subharmonic in the entire complex plane \cite{CS} and bounded from
below,
\begin{equation}\label{lb}
\gamma (z) \ge \frac{1}{2} E\log|a_1/c_1|\hspace{3ex} \hbox{for all $z$.}
\end{equation}
This inequality easily follows from  $\det S_n(z)=\prod_{j=1}^n
a_j/c_j$. The subharmonicity implies that $\Delta \gamma $, where
$\Delta$ is the distributional Laplacian in variables $\re z$ and
$\im z$, defines a measure on $\C$, see e.g. \cite{H1}. Our main
result is as follows.

\bigskip
\begin{Thm} \label{thm1} Let $\mu_n$ be the normalized eigenvalue counting measure of
$J_n$, i.e. $ \mu_n=\frac{1}{n}\sum_{l=1}^{n}\delta_{z_l}$, where
$z_1, \ldots z_n$ are the eigenvalues of $J_n$. Then:
\begin{itemize}
\item[(a)]
With probability one, $\mu_n$ converges weakly to
$\mu=\frac{1}{2\pi}\Delta \gamma$ as $n\to\infty$.
\item[(b)] (Thouless formula) For every $z\in \C$
\begin{equation}\label{thouless}
\gamma(z)=\int_{\C} \log|w-z|\ d\mu (w)-E\log|c_1|.
\end{equation}
\item[(c)] The limiting eigenvalue counting measure $\mu$ is log-H\"older
continuous. More precisely, for any
$B_{z_0,\delta}=\{z:\hspace{0.5ex} |z-z_0|\le \delta\}$, $0<\delta
<1$,
\begin{equation}\label{hol}
\mu(B_{z_0,\delta})\le \frac{C(z_0,\delta)}{\log\frac{1}{\delta}},
\end{equation}
where $C(z_0,\delta)\to 0$ as $\delta \to 0$.
\end{itemize}
\end{Thm}
\bigskip

We deduce Theorem \ref{thm1} from Theorem \ref{thm1a} which is of
independent interest in the context of second order difference
equations.

\bigskip
\begin{Thm}\label{thm1a}
With probability one
\begin{equation}\label{g}
\lim_{n\to\infty}\frac{1}{n}\log |f_{n+1}(z)|=\gamma(z)
\end{equation}
for almost all $z$ with respect to the Lebesgue measure on $\C$.
\end{Thm}

\bigskip

To prove Theorem \ref{thm1a}, we use the theory of products of
random matrices. Of course this is unnecessary in the Hermitian
case. In this case (\ref{g}) and the Thouless formula
(\ref{thouless}) follow directly from the fact that $\mu_n$
converges weakly to a limiting measure $\mu$ \cite{AS,FP} and the
latter can be established independently and by more elementary
means. We would like to emphasize that in the non-Hermitian case
we follow the opposite direction route: the weak convergence of
$\mu_n$ and the Thouless formula are deduced from (\ref{g}). To
this end we make use of the relation
\begin{equation}\label{poisson}
\mu_n = \frac{1}{2\pi n}\Delta \log |f_{n+1}|,
\end{equation}
where the equality is to be understood in the sense of
distribution theory. Relation (\ref{poisson}) is well known in the
function theory. It holds for arbitrary polynomial of degree $n$
and can be easily derived with the help of the Gauss-Green
formula. In this general setup it was shown by Widom \cite{W1,W2}
that if the measures $\mu_n$ for all $n$ are supported inside a
bounded region and in the limit $n\to\infty$ the function
$p_n(z)=\int_{\C} \log |z-w| d\mu_n(w)$ converges to a limiting
function $p(z)$ almost everywhere in the complex plane then
$\mu_n$ converges weakly to $\mu=\frac{1}{2\pi}\Delta p$. We shall
need the following simple extension of this result to the case
when the supports of $\mu_n$ are not necessarily bounded.

Let  $A_n$ be a (deterministic) sequence of square matrices of
increasing dimension $n$, and
\[
p_n(z)=\frac{1}{n}\log |\det (A_n - zI_n)|=\int_{\C}\log|w-z|
d\mu_n(w),
\]
where $I_n$ is $n\times n$ identity matrix and $\mu_n=
\frac{1}{2\pi}\Delta p_n$ is the normalized eigenvalue counting
measure of $A_n$. Define
\begin{equation}\label{a4}
\tau_R=\limsup_{n\to\infty}\int_{|w|\ge R} \log |w|\ d\mu_n(w),
\hspace{3ex} R\ge 1.
\end{equation}

\bigskip

\begin{Prop}\label{1:prop1}  Assume that there is a function
$p\colon \C \to [-\infty, +\infty)$ such that $p_n(z)\to p(z)$ as
$n\to\infty$ almost everywhere in $\C$. If $\tau_1<+\infty$ then
it follows that $p$ is locally integrable, $\mu=\frac{1}{2\pi}
\Delta p $ is a unit mass measure,
\begin{equation}\label{a8}
\int_{|w|\ge 1} \log |w|\ d\mu(w) \le \tau_1 <+\infty,
\end{equation}
and the sequence of measures $\mu_n$ converges weakly to $\mu$ as
$n\to\infty$. If, in addition, $\lim_{R\to\infty}\tau_R=0$ then we
also have that
\begin{equation}\label{log}
p(z)=\int_{\C} \log|w-z| d\mu(w).
\end{equation}

\end{Prop}

\bigskip

\emph{Remark.} In view of (\ref{a8}), the integral on the RHS in
(\ref{log}) is a locally integrable function of $z$ taking values
in $[-\infty, +\infty)$.

For the sake of completeness, we give a proof of this Proposition
in Appendix \ref{appa}.

In order to estimate the tails of eigenvalue distributions as
required in the above Proposition \ref{1:prop1} we use the
following inequalities\footnote{Note that
$\log\det(I_n+A_nA_n^*)=\tr \log (I_n+A_nA_n^*)$.}:
\begin{equation}\label{tau1}
\tau_1 \le\limsup_{n\to\infty} \frac{1}{2n}\log \det
(I_n+A_nA_n^*),
\end{equation}
and for any $R>1$ and $\delta >0$
\begin{equation}\label{taur}
\tau_R \le \frac{1}{\log^{\delta} R }\limsup_{n\to\infty}
\frac{1}{2^{1+\delta}n}\tr \log^{1+\delta} (I_n+A_nA_n^*).
\end{equation}
These inequalities can be derived with the help of Weyl's Majorant
Theorem, for details of derivation see Appendix \ref{appb}.

Let us now return to the random Jacobi matrices $J_n$.
Straightforward but tedious calculations show\footnote{For any
Hermitian matrix $H=||H_{jk}||_{j,k=1}^n$ we have $H\le
D=\diag(d_1, \ldots, d_n)$ with $d_j=\sum_{k=1}^n |H_{jk}|$,
$j=1,\ldots ,n$. Therefore if $f$ is a nondecreasing function
then, by the Courant-Fisher minimax principle, $\tr \log f(H) \le
\tr f(D)=\sum_{j=1}^n f(d_j)$.} that
\[
\frac{1}{n}\tr \log^{1+\delta} (I_n+J_nJ_n^*)\le \frac{\alpha}{n}
\sum_{j=1}^n\log^{1+\delta} (1+\beta |\mathbf{v}_j|^2),
\hspace{1ex} \hbox{where $\mathbf{v}_j= (a_j,b_j,c_j) $},
\]
for some $\alpha,\beta>0$ independent of $\mathbf{v}_j$'s and $n$.
Therefore if the random sequence $\mathbf{v}_j$ is stationary and
\begin{equation}\label{m}
E\log^{1+\delta} [1+|\mathbf{v}_1|^2] <\infty \hspace{3ex}
\hbox{for some $\delta >0$}
\end{equation}
then the Ergodic Theorem asserts that with probability one the
limits in (\ref{tau1}) (\ref{taur}) are finite which implies
$\tau_1<\infty$ and $\lim_{R\to\infty} \tau_R=0$, as required in
Proposition \ref{1:prop1}. The assumptions of stationarity and
(\ref{m}) are less restrictive than assumptions A1-A3. However we
are only able to prove Theorem \ref{thm1a} (which is the main
ingredient to our proof of Theorem \ref{thm1}) under these more
restrictive assumptions.

\section{Products of random matrices }\label{sec2}

Our proof of Theorem \ref{thm1a} makes use of several facts from
the theory of products of random $2\times2$ matrices. We list
these facts below (Propositions \ref{prop1} - \ref{prop3}).

Let $\nu$ be a probability distribution on the group $Gl(2,\C)$ of
invertible complex $2\times 2$ matrices and $g_k$ be an infinite
sequence of independent samples from this distribution.

As before $S_n=g_n\cdot \ldots \cdot g_1$ for $n=1,2,\ldots.$ By
$P(\C^2)$ we denote the projective space on which every
non-degenerate matrix $g$ acts in a natural way. Let $\kappa$ be a
probability measure on $P(\C^2)$. We say that $g$ preserves
$\kappa$ if $\kappa(g^{-1}.B) =\kappa(B)$ for any Borel set $B$
(here $g.x$ is the result of the action of $g$ on $x\in P(\C^2)$).
By $G_{\nu}$ we denote the closure of the subgroup of $Gl(2,\C)$
generated by all matrices belonging to the support of $\nu$. We
say that $G_{\nu}$ preserves $\kappa$ if $\kappa$ is preserved by
every $g\in G_{\nu}$.

\bigskip

\begin{Prop}\label{prop1}
Let $\lambda_1^{(n)}\ge \lambda_2^{(n)}$ be
the singular values of $S_n$.
If
\begin{equation}\label{cond1}
\mbox{$E\log ||g||$ and $E\log |\det
g|$ are both finite}
\end{equation}
then with probability one
the following limits
\begin{equation}\label{4}
\lim_{n\to\infty}\frac{1}{n}\log
\lambda_j^{(n)}=\gamma_j,\hspace{3ex}  j=1,2,
\end{equation}
exist and are nonrandom.
\end{Prop}

\bigskip

The limiting values $\gamma_1$ and $\gamma_2$ are called the
Lyapunov exponents of the sequence $S_n$.

\bigskip

\begin{Prop}\label{prop2} If in addition to condition
(\ref{cond1}), no measure $\kappa$ is preserved by $G_{\nu}$ then
the Lyapunov exponents of the sequence $S_n$ are distinct, i.e.
$\gamma_1>\gamma_2$.
\end{Prop}

\bigskip

\begin{Prop}\label{prop3} If condition (\ref{cond1}) is satisfied
and no measure $\kappa$ is preserved by $G_{\nu}$ then
\begin{itemize}
\item[(i)]
For any unit vector $x$ the probability is one that
\begin{equation}\label{5}
\lim_{n\to\infty}\frac{1}{n}\log ||S_n  x || =\gamma_1.
\end{equation}

\item[(ii)] If in addition
$E(||g||^{\delta}+||g^{-1}||^{\delta})<\infty$ for some $\delta >
0$ then for any positive $\varepsilon$ there is a constant $\rho
(\varepsilon)>0$ such that uniformly in $x,\ ||x||=1,$
\begin{equation}\label{6}
Prob\left(\left|\log ||S_n x|| -n\gamma_1 \right|\ge \varepsilon n
|\right) \le e^{-n\rho (\varepsilon)}.
\end{equation}
\end{itemize}
\end{Prop}

\bigskip

\emph{Remarks.} 1. As all norms in $\C^2$ are equivalent,
the choice
of norm in (\ref{5}) and (\ref{6}) is not important. However it is
convenient to deal with the standard Euclidian norm.

2. Propositions \ref{prop1} - \ref{prop3} are well known in the
classical case of the real matrices,
see e.g. \cite{O, BL} for proofs of Propositions \ref{prop1}
and \ref{prop3} and \cite{F,V} for proofs of Proposition
\ref{prop2}. For complex matrices,
Propositions \ref{prop1} and \ref{prop3} are proved in the
same way as in \cite{O, BL}.  However, the proof of
Proposition \ref{prop2} is somewhat
different from that given in \cite{F,V}.
We shall now discuss the necessary changes which would allow
the interested reader to reconstruct the proof in question simply by
examining the one in \cite{V}. Namely, the main ingredient of
this proof is the fact that the mapping
$g\mapsto T_g$, where
\[
\left(T_gf\right)(x)=f(g^{-1}x)||g^{-1}x||^{-\frac{m}{2}},
\]
defines a unitary representation of the group $SL(m,\R)$ in
Hilbert space $L_2(\mathcal{S}_m,dl)$ with $dl$ being the natural Lebesgue
measure on the unit sphere $\mathcal{S}_m\in \R^m$.
(Obviously, we are interested in the case when $m=2$.)

In the case of the complex
space the representation is defined by
\[
\left(T_gf\right)(x)=f(g^{-1}x)||g^{-1}x||^{-m},
\]
in Hilbert space $L_2(\mathcal{S}_m,dl)$ with $dl$ being again the natural Lebesgue
measure on the unit sphere $\mathcal{S}_m\in \C^m$.
After that the proof proceeds in the way suggested in \cite{V}.

\section{Proofs of Theorems \ref{thm1} and \ref{thm1a}}\label{sec3}

In order to be able to apply Propositions \ref{prop1} --
\ref{prop3} we have to verify that under assumptions A1--A3 our
matrices $g_j$ defined in (\ref{ggg}) satisfy the conditions of
these Propositions.

Is is apparent that assumption A2 guarantees that condition
(\ref{cond1}) is satisfied and
$E(||g||^{\delta}+||g^{-1}||^{\delta})<\infty$. It remains to
check that assumption A3 implies that no measure $\kappa$ is
preserved by $G_{\nu}$ (here $\nu$ is the measure induced on the
group of matrices by the distribution of $(a_1,b_1,c_1)$). To this
end we note that if
\[
g=
\begin{pmatrix}
\frac{z-b}{c}  & \frac{-a}{c}    \\1   &  0    \\
\end{pmatrix}\hspace{1ex} \hbox{and}\hspace{1ex}
g^{\prime}=
\begin{pmatrix}
\frac{z-b^{\prime}}{c^{\prime}}  & \frac{-a }{c^{\prime}}   \\1   &  0    \\
\end{pmatrix}
\]
then
\[
g{g^{\prime}}^{-1}=
\begin{pmatrix}
 \frac{c^{\prime}a}{a^{\prime}c} &
  \frac{z-b}{c}-\frac{(z-b^{\prime})a}{a^\prime c}   \\0   &  1    \\
\end{pmatrix} \hspace{1ex} \hbox{and}\hspace{1ex}
{g^{\prime}}^{-1}g=\begin{pmatrix}
 1 &
  0  \\\frac{z-b^{\prime}}{a^\prime}-\frac{(z-b)c^\prime}{ca^\prime} &
  \frac{c^{\prime}a}{a^{\prime}c}   \\
\end{pmatrix} \hspace{1ex} \hbox{and}\hspace{1ex}
\]
It remains to check that for almost all $z$ that the group $G$
generated by the matrices $g,\ g^{\prime}$ is rich enough in the
sense that no measure is preserved by all matrices of this group.
The main idea is as follows. For a "typical" $z$ we construct two
matrices, say $B$ and $D$, from $G$ such that the eigenvalues of
$B$ are of different moduli. It is easy to see then that the only
measure preserved by all matrices of the form $B^n,\
-\infty<n<\infty $ is the one supported by the lines in $P(\C^2)$
generated by the eigenvectors of $B$. The matrix $D\in G$ is then
chosen so that its action on $P(\C^2)$ does not preserve these
lines which means that the measure in question does not exist. We
would like to emphasize that the presence of the parameter $z$
plays a crucial role in this situation.

More precisely, if $z$ is such that
\[
2\arg (z-b)\not=\arg(ac)
\]
then the matrix $g$ has eigenvalues with different moduli. In other words
the moduli are different if $z$ does not belong to a certain half line.
The  $g^\prime$
plays then the role of $D$ (once again when $z$ lies outside of certain
curves). This statement can be checked by direct calculation and is
sufficient for our purposes.

However, in some important cases much more precise statements can be made.
In particular if $\frac{c^{\prime}a}{a^{\prime}c}=1$ then each of triangular
 matrices  $(g{g^{\prime}}^{-1}) $ and ${g^{\prime}}^{-1}g$ is non-trivial
 for all but may be two values of $z$ and a similar idea  applies,
see \cite{BL} page 213.

Now we are in a position to apply Propositions \ref{prop1}
- \ref{prop3}.
For any two non-zero vectors $x$ and $y$ define
\[
d(x, y)=\sqrt{1-\frac{|(x,y)|^2} {(x,x)(y,y)}},
\]
where $(\cdot,\cdot)$ is the scalar product in $\C^2$. The
function $d(x,y)$ is the natural angular distance between $x$ and
$y$ on the projective space $P(\C^2)$.

The following Lemma is the key element in the proof of Theorem
\ref{thm1a}. (In this Lemma and thereafter the abbreviation a.s.
refers to the probability measure, i.e. any equality with the
letters a.s. above it holds with probability one)

\begin{Lemma} \label{lemma1} Suppose that the conditions of Propositions
\ref{prop1} - \ref{prop3} are satisfied. If $y_n$ is a sequence of
random unit vectors in $\C^2$ such that
\begin{equation}\label{7}
||S_n y_n||=e^{n\gamma_2 +\epsilon_n}, \hspace{1ex}
\hbox{where}\hspace{1ex} \epsilon_n \overset{\rm a.s.} {=} o(n)
\hspace{1ex} \hbox{as}\hspace{1ex} n\to\infty,
\end{equation}
then for any fixed unit vector $x$ and any $\delta
>0$ there is a constant $r(x,\delta)>0$ such that
\begin{equation}\label{8}
Prob\left\{d(x,y_n)\le e^{-n\delta}\right\} \le e^{-n r (x,
\delta)}
\end{equation}
for all sufficiently large $n$.
\end{Lemma}

\smallskip

\noindent \emph{Proof.} For any $n$ one can always find two
orthogonal unit vectors $u_n$ and $v_n$ such that
$S_n^{*}S_nu_n=\lambda_1^{(n)}$ and
$S_n^{*}S_nv_n=\lambda_2^{(n)}$. In view of Proposition
\ref{prop1},
\[
||S_n u_n||=e^{n\gamma_1+\epsilon_n^{\prime}}\hspace{1ex}
\hbox{and}\hspace{1ex} ||S_n
v_n||=e^{n\gamma_2+\epsilon_n^{\prime\prime}}, \hspace{1ex}
\hbox{where}\hspace{1ex} \epsilon_n^{\prime},
\epsilon_n^{\prime\prime} \overset{\rm a.s.} {=} o(n).
\]
Obviously the sequence $v_n$ satisfies condition (\ref{7}) and we
first prove the large deviation estimate (\ref{8}) for this
sequence.

Let $x$ be a fixed unit vector. Then $x=(x,u_n)u_n+(x,v_n)v_n$ for
every $n$, and, since $|(x,u_n)|=d(x,v_n)$ and $|(x,v_n)|\le 1$,
we have that
\[
||S_nx|| \le  d(x,v_n)||S_nu_n|| + ||S_nv_n||.
\]
Therefore if $d(x,v_n) \le e^{-n\delta}$ then
\[
\log ||S_nx|| \le n\gamma_1 + \log (e^{-n\delta
+\epsilon_n^{\prime}}
+e^{-n(\gamma_1-\gamma_2)+\epsilon_n^{\prime\prime}}),
\]
and hence with probability one,
\[
\log ||S_nx|| - n\gamma_1 \le -n\min(\delta, \gamma_1-\gamma_2)
+o(n).
\]
It follows now from Proposition \ref{prop3} that
\begin{equation}\label{9}
Prob\left\{d(x,v_n)\le e^{-n\delta}\right\} \le e^{-n r (x,
\delta)}
\end{equation}
for some $r (x, \delta)$ and all $n>n_0$ where $n_0$ depends on
the matrices $S_n$, and also on $x$ and $\delta$.

Now, let $y_n$ be an arbitrary sequence of random unit vectors
satisfying condition (\ref{7}), and  let $y_n^{\bot}$ be a
sequence of unit vectors orthogonal to $y_n$, i.e.
$(y_n,y_n^{\bot})=0$ for all $n$. Obviously, $d(u_n,
y_n^{\bot})=|(u_n,y_n)|$ and, since $S_n^{*}S_nu_n=e^{2
n\gamma_1+2\epsilon_n^{\prime}}u_n$, we have that with probability
one
\[
d(u_n, y_n^{\bot}) = e^{-2 n\gamma_1+o(n)}|(S_nu_n, S_ny_n)|\le
e^{-n(\gamma_1-\gamma_2)+o(n)}.
\]
It is then apparent that $d(v_n, y_n)$ is also exponentially small
for large $n$ and therefore the large deviation estimate (\ref{8})
for $y_n$ follows from (\ref{9}). \hfill $\Box$

\bigskip

\emph{Proof of Theorem \ref{thm1a}.} Let
\[
x=\left(
\begin{array}{l}
0 \\ 1 \\
\end{array}
\right) \hspace{3ex} \hbox{and}\hspace{3ex} y_n=\left(
\begin{array}{l}
f_{n+1}(z) \\ f_n(z) \\
\end{array}
\right), \hspace{1ex} n=1,2,\ldots .
\]
Then
\[
d^2(x,y_n)=\frac{|f_{n+1}(z)|^2}{|f_{n+1}(z)|^2+|f_{n}(z)|^2}
=\frac{|f_{n+1}(z)|^2}{||y_n||^2},
\]
and therefore
\begin{equation}\label{10}
\frac{1}{n}\log |f_{n+1}(z)|=\frac{1}{n}\log d(x,y_n)
+\frac{1}{n}\log ||y_n||.
\end{equation}
In view of (\ref{3a}) and Proposition \ref{prop3}(i),
\begin{equation}\label{11}
\lim_{n\to\infty}\frac{1}{n}\log ||y_n||\overset{\rm a.s.} {=}
\gamma_1(z),
\end{equation}
where $\gamma(z)$ is the upper Lyapunov exponent of the sequence
of transfer matrices $S_n(z)$. On the other hand, $S^{-1}_n(z)y_n
=(1,0)^{T}$, and therefore
\[ \lim_{n\to\infty}\frac{1}{n}\log
\frac{||S^{-1}_n(z)y_n||}{||y_n||}\overset{\rm a.s.}
{=}-\gamma_1(z).
\]
It follows now from Lemma \ref{lemma1} (applied to the matrices
$S_n^{-1}(z)$ and the vectors $x$ and $y_n/||y_n||$\footnote{If
$\gamma_1$ and $\gamma_2$ are the Lyapunov exponents of a sequence
$S_n$ then the sequence $S_n^{-1}$ has the Lyapunov exponents
$-\gamma_2$ and $-\gamma_1$.}) and the Borel-Cantelli Lemma that
\[
\lim_{n\to\infty}\frac{1}{n}\log d(x,y_n) \overset{\rm a.s.} {=}
0.
\]
Therefore, in view of (\ref{10}) and (\ref{11}), for any fixed $z$
the probability is one that
\begin{equation}\label{12}
\lim_{n\to\infty}\frac{1}{n}\log |f_{n+1}(z)|=\gamma_1(z).
\end{equation}
But then the probability is one that (\ref{12}) holds almost
everywhere in the complex plane. This follows from the Fubini
Theorem. Our proof of Theorem \ref{thm1a} is complete.

\bigskip

\emph{Proof of Theorem \ref{thm1}.} As explained in introduction,
under assumptions A1 -- A3, the probability is one that $\tau_1\le
C$ for some non-random $C <+\infty$ and
$\lim_{R\to\infty}\tau_R=0$. Therefore parts (a) and (b) of
Theorem \ref{thm1} follow immediately from Theorem \ref{thm1a} by
the way of Proposition \ref{1:prop1}.

The log-H\"older continuity of $\mu$ is a corollary of the
Thouless formula and the fact that the Lyapunov exponent $\gamma
(z)$ is bounded from below. This is very much in the same way as
in the Hermitian case, see \cite{CS}.

To prove (\ref{hol}), we first note that the integral
$\int_{\C}\log |w-z|d\mu (w)$ converges absolutely for every $z$.
Indeed, it follows from (\ref{a8}) that
\[
\int_{|w-z|\ge 1} \log |w-z| d\mu(w) <+\infty,
\]
and this inequality together with the Thouless formula and the
lower bound (\ref{lb}) imply that
\[
\int_{|w-z| \le 1} |\log |w-z|| d\mu(w) <+\infty
\]
as well. Therefore,
\begin{equation}\label{ho}
C(z,\delta) := \int_{|w-z|\le \delta}|\log|w-z|| d\mu(w) \to 0
\hspace{3ex} \hbox{as $\delta \to 0$.}
\end{equation}
Obviously, for $\delta <1$,
\[
C(z,\delta)=\int_{|w-z|\le \delta}\log(1/|w-z|) d\mu(w) \ge \frac{
\mu(B_{z,\delta})}{\log(1/\delta)},
\]
and part (c) of Theorem \ref{thm1} follows. Our proof of Theorem
\ref{thm1} is now complete.

\bigskip

\appendix

\section{Appendix} \label{appa}
\emph{Proof of Proposition \ref{1:prop1}.} The local integrability
of $\log|z|$ and the condition $\tau_1<+\infty$ imply that the
functions $p_n(z)$ are uniformly integrable on bounded sets in
$\C$. It follows from this that $p(z)$ is locally integrable and
$p_n\to p$ as $n\to\infty$ in $D^{\prime}(\C)$, the space of
Schwartz distributions in $\C$. Since $\Delta$ is continuous on
distributions, we also have that $\Delta p_n \to \Delta p$ in
$D^{\prime}(\C)$. Obviously $\Delta p \ge 0$, hence $\Delta p$ is
defined by a measure, see e.g. \cite{H2}. As any sequence of
measures converging as distributions must converge weakly we
conclude that $\mu_n=\frac{1}{2\pi}\Delta p_n \to
\mu=\frac{1}{2\pi}\Delta p$ weakly as measures.

For any $R>1$,
\[
\int_{|w|\ge R} d\mu_n(w)\le\frac{1}{\log|R|}\ \int_{|w|\ge 1}
\log |w| d\mu_n(w).
\]
Therefore the inequality $\tau_1 <+\infty$ implies that the
sequence of measures $\mu_n$ is tight, and hence cannot lose mass.
As each of $\mu_n$ has unit mass, so has the limiting measure
$\mu$.

It follows from the weak convergence of $\mu_n$ to $\mu$ and
(\ref{a4}) that
\[
\int_{1\le |w|\le R} \log |w| d\mu (w) \le \lim_{n\to\infty}
\int_{1\le |w|\le 2R} \log |w| d\mu_n (w) \le \tau_1
\]
for any $R>1$. This implies (\ref{a8}). Similarly, if
$\lim_{R\to\infty}\tau_R=0$ then
\begin{equation}\label{a8a}
\lim_{R\to\infty} \int_{|w|\ge R} \log |w| d\mu (w)=0.
\end{equation}

It remains to prove relation (\ref{log}). It will suffice to show
that
\begin{equation}\label{ad}
p_n \to\ \int_{\C}\log|w-\cdot|d\mu(w) \hspace{3ex}\hbox{in
$D^{\prime}(\C)$}
\end{equation}
when $n\to\infty$. Let $\psi (z)$ be a continuous function with
bounded support. Then
\[
\int_{\C}p_n(z)\psi(z)d^2z=\int_{\C}g(w)d\mu_n(w)
\]
with
\[
g(w)=\int_{\C}\psi(z) \log |w-z| d^2z.
\]
The function $g$ is continuous and $g(w)=O(\log|w|)$ when
$|w|\to\infty$. Assume now that $\lim_{R\to\infty}\tau_R=0$. Then
\[
\lim_{R\to\infty} \limsup_{n\to\infty} \int_{|w|\ge R} |g(w)|
d\mu_n(w) =0,
\]
and
\[
\lim_{R\to\infty} \int_{|w|\ge R} |g(w)| d\mu (w) =0
\]
because of (\ref{a8a}). It now follows from the weak convergence
of $\mu_n$ to $\mu$ that
\[
\lim_{n\to\infty}\int_{\C}g(w)d\mu_n(w)=\int_{\C}g(w)d\mu (w).
\]
Therefore
\[
\lim_{n\to\infty}\int_{\C}p_n(z)\psi(z)d^2z =\int_{\C}g(w)d\mu
(w)=\int_{C} \left\{ \int_{\C}\log|w-z|d\mu(z)\right\} \psi(z)
d^2z,
\]
and (\ref{ad}) follows.

\section{Appendix} \label{appb}

\emph{Derivation of inequalities (\ref{tau1}) and (\ref{taur}).}
Let $z_1, \ldots , z_n$ and $s_1, \ldots , s_n$ be respectively
the eigenvalues and singular values of $A_n$ labeled so that
$|z_1|\ge |z_2| \ge \ldots \ge |z_n|$ and $s_1\ge s_2 \ge \ldots
\ge s_n$. Weyl's Majorant Theorem, see \cite{GohK}, page 39,
asserts that
\[
\sum_{j=1}^m F(|z_j|) \le \sum_{j=1}^m F(s_j), \hspace{3ex}
m=1,2,\ldots, n,
\]
for any function $F(t)$ $(0\le t<\infty)$  such that $F(e^x)$ is
convex on $\R$. Obviously the function $\log^{1+\delta}(t)$
satisfies this requirement for $\delta \ge 0$, and therefore
\[
\int_{|w|\ge 1}\log^{1+\delta} |w|
d\mu_n(w)=\frac{1}{n}\sum_{|z_j|\ge 1}\log^{1+\delta} |z_j|\le
\frac{1}{n}\sum_{|z_j|\ge 1}\log^{1+\delta} |z_j| \le
\frac{1}{n}\sum_{j=1}^m\log^{1+\delta} s_j
\]
where $m$ is the number of eigenvalues of $A_n$ such that
$|z_j|\ge 1$. Obviously,
\[
\sum_{j=1}^m\log
s_j=\frac{1}{2^{1+\delta}}\sum_{j=1}^m\log^{1+\delta} (s_j^2) \le
\frac{1}{2^{1+\delta}}\sum_{j=1}^n\log^{1+\delta}
(1+s_j^2)=\frac{1}{2^{1+\delta}}\tr \log^{1+\delta}
(I_n+A_nA_n^*),
\]
and therefore
\[
\int_{|w|\ge 1}\log^{1+\delta} |w| d\mu_n(w) \le
\frac{1}{2^{1+\delta}n} \tr \log^{1+\delta} (I_n+A_nA_n^*),
\hspace{3ex} \delta \ge 0,
\]
which implies (\ref{tau1}) and (\ref{taur}).

\end{document}